\documentclass[%
 reprint,
 amsmath,amssymb,
 aps,
]{revtex4-2}

\usepackage{graphicx}
\usepackage{mathtools}
\usepackage{dcolumn}
\usepackage{bm}
\usepackage{braket}
\usepackage{graphics}
\usepackage{color}

\makeatletter
\def\Dated@name{Fecha: }
\makeatother

\begin{document}

\preprint{APS/123-QED}

\title{Dispersive $\pi\pi\rightarrow K\bar K$ amplitude and giant CP violation in $B$ to three light-meson decays at LHCb}

\author{R. \'Alvarez Garrote}
 \altaffiliation[]{}
\affiliation{%
Centro   de   Investigaciones   Energéticas   Medioambientales   y   Tecnológicas   (CIEMAT), Madrid, SPAIN.
}%
\author{J. Cuervo}
 \altaffiliation[]{}
\affiliation{%
 Departamento de Física Teórica. Universidad Complutense. 28040 Madrid. SPAIN.
}%
\author{P. C. Magalh\~aes, J. R. Pel\'aez}
 \email{contact author:jrpelaez@ucm.es}
 \altaffiliation[]{}
\affiliation{%
 Departamento de Física Teórica. Universidad Complutense and IPARCOS. 28040 Madrid. SPAIN.
}%


\begin{abstract}
The LHCb collaboration has recently reported  the largest CP violation effect from a single amplitude, as well as other giant CP asymmetries in several  $B$-meson decays into three charmless light mesons. It is also claimed that this is predominantly due to $\pi\pi\rightarrow K\bar K$
rescattering in the final state, particularly in the 1 to 1.5 GeV region. In these analyses the $\pi\pi\rightarrow K\bar K$ amplitude is by default estimated from the $\pi\pi$ elastic scattering amplitude and does not describe the existing $\pi\pi\rightarrow K\bar K$ scattering data.
Here we show how the recent model-independent dispersive analysis of  $\pi\pi\rightarrow K\bar K$ data can be easily implemented in the LHCb formalism.
This leads to a more accurate description of the asymmetry,  while being consistent with the measured scattering amplitude and confirming the prominent role of hadronic final state interactions, paving the way for more elaborated  analyses.
\end{abstract}

\maketitle


In a series of recent works the LHCb Collaboration has reported the observation of direct CP symmetry violation (CPV) in charged $B$-meson charmless decays into three pseudoscalar mesons. The relevance of these processes is that the observation of CPV requires the interference between a ``weak" phase, which changes sign for opposite CP states, with a CP invariant ``strong" phase. While the source of the first one is well understood from the Cabibbo-Kobayashi-Maskawa matrix \cite{Cabibbo:1963yz,Kobayashi:1973fv} of the Standard Model, and can be easily calculated in standard perturbation theory, the latter is much more troublesome due to its strong character. Moreover, it is long thought that it can be produced from short-distance quark-level contributions and/or long-distance hadronic final state interactions (FSI) \cite{Wolfenstein:1990ks,Suzuki:1999uc,Suzuki:2007je}. The relevance of CPV in three-body decays (see \cite{Bediaga:2020qxg} for a recent review) is that it can be studied, not only from the total or integrated charge asymmetry, which is  a single number, but from the phase-space distribution of the decay, which  is a function that depends on two energy variables and is much richer in structure. Moreover, the rescattering of final state hadrons is dominated by resonances that can yield huge variations throughout the phase-space distributions. The energy dependence of these distributions may allow disentangling different sources of strong phases in CPV.

In particular, CPV both in the local and integrated phase-space asymmetries between the opposite charge  $B^\pm\rightarrow K^\pm \pi^+\pi^-$ and $B^\pm\rightarrow K^\pm K^+K^-$ decays was first observed at LHCb in 2013 \cite{Aaij:2013sfa}, followed by the observation of the corresponding asymmetries in  $B^\pm\rightarrow \pi^\pm \pi^+\pi^-$ and $B^\pm\rightarrow \pi^\pm K^+K^-$ \cite{Aaij:2013bla}. These $B\rightarrow 3M$ analyses were soon superseded with larger statistical samples in 2014 \cite{Aaij:2014iva}. 
Whereas integrated asymmetries came up of the order of 2 to 12 \%, it was shown that local asymmetries could be very large,  when looking at localized regions in the Dalitz plots. The collaboration suggested that FSI may be a determinant factor for this giant CP violation.  In particular, asymmetries became very large when the Dalitz plot was projected on the invariant mass of the opposite-charged final mesons, and in the 1 to 1.5 GeV range, which was  associated to the  inelastic isoscalar $S$-wave $\pi^+\pi^-\leftrightarrow K^+ K^-$ FSI. Even more accurate CPV results have just been presented \cite{LHCb:2022nyw}, still supporting the relevance of FSI, which could also be important for CPV in charm decays \cite{Bediaga:2022sxw}.

It is only very recently that the LHCb has performed the full amplitude analyses of their run I data on $B^\pm\rightarrow \pi^\pm K^+K^-$ \cite{Aaij:2019qps} and $B^+\rightarrow \pi^+ \pi^+\pi^-$ \cite{Aaij:2019hzr,Aaij:2019jaq}.  Their most striking feature is that, for $B^\pm\rightarrow \pi^\pm K^+K^-$, the collaboration claims $\pi\pi\rightarrow K\bar K$ S-wave rescattering to have {\em ``the largest CP asymmetry reported to date for a single amplitude of $(-66\pm 4 \pm 2)\%$"}. For $B^+\rightarrow \pi^+ \pi^+\pi^-$  a similarly large value $\sim 45 \%$, is also found.

The inelastic FSI framework for CPV dates back to Wolfenstein and Suzuki in \cite{Wolfenstein:1990ks,Suzuki:1999uc,Suzuki:2007je}. The LHCb amplitude analyses \cite{Aaij:2019qps,Aaij:2019hzr,Aaij:2019jaq} used a very appealing particularization to $B\rightarrow3M$ in \cite{Bediaga:2013ela,Nogueira:2015tsa} 
 (Other models are also used for $B\to3\pi$ in \cite{Aaij:2019hzr,Aaij:2019jaq}).  This model is relevant in the 1 to 1.5 GeV region, where final-state multiplicity is low and the CPT
constraint is more enlightening.
It also assumes that only two particles re-scatter whereas the third is a spectator.
In this formulation,  the $\pi^+\pi^-\leftrightarrow K^+ K^-$ complex isoscalar partial $S$-wave should be described by its modulus and phase $\delta_{\pi\pi K\bar K}$. However, 
in the \cite{Bediaga:2013ela,Nogueira:2015tsa} formalism  and its implementation by LHCb \cite{Aaij:2019qps,Aaij:2019hzr,Aaij:2019jaq},
or modification by \cite{Cheng:2016shb,Cheng:2020ipp}, 
the $\pi\pi\to K\bar K$ interaction is not used.
Instead, it is assumed that $K\bar K$ and $\pi\pi$ are the only available states and, in addition, the phase is crudely estimated as $\delta_{\pi\pi K K}\sim2 \delta_{\pi\pi\pi\pi}$ whereas its elasticity is obtained 
from that of $\pi\pi$ scattering. Of course, 
in this way they could use the model-independent dispersive analysis of $\pi\pi$ scattering data in \cite{Pelaez:2004vs}. It is true that meson-meson scattering experiments are plagued with systematic errors and have been usually described with crude models (see \cite{Pelaez:2015qba,Pelaez:2020gnd} for reviews). Model-independent parameterizations can only be obtained through dispersive methods, whose relevance has been repeatedly emphasized in the context of heavy particle hadronic decays \cite{Bigi:2013aca,Bigi:2014xna}. However, as we will show below this estimate does not reproduce the $\pi\pi\to K\bar K$ data.
Moreover, it violates Watson's Theorem \cite{Watson:1952ji}, which implies that at $KK$ threshold, and for partial waves with given total angular momentum and isospin, $\delta_{\pi\pi KK}=\delta_{\pi\pi\pi\pi}$, without that factor of 2.
Furthermore, the poorly known $\pi\pi$ elasticity and the factor of 2 amplifying the already large  $\delta_{\pi\pi\pi\pi}$ error gives rise to huge uncertainties in the description of the asymmetry FSI. Despite this treatment may provide a hint of the relevance of FSI, it definitely calls for an implementation using the realistic $\pi\pi\to K \bar K$ amplitude, consistent with data and fundamental constraints.

Fortunately, a dispersively constrained $\pi\pi \to K \bar K$ data analysis has become recently available \cite{Pelaez:2018qny,Pelaez:2020gnd}. It provides precise and model-independent parameterizations of phases and moduli for several partial waves, including the isoscalar $S$-wave. Here we show how to implement easily this dispersive parameterization within the formalism presently used by LHCb, proposed in \cite{Bediaga:2013ela,Nogueira:2015tsa}, and how it improves dramatically the accuracy of the FSI contribution to these CPV asymmetries. Moreover, it unveils hadronic structures that were masked in the uncertainties, while providing a sound support for the FSI prominent role in these giant CP violations. Implementing these amplitudes in future LHCb  analyses will provide much more precise descriptions and may allow to understand further
hadronic details otherwise swamped by the huge uncertainties of the present estimates.

Let us briefly recall the FSI formalism in \cite{Bediaga:2013ela,Nogueira:2015tsa}, with simplified notation and assuming $CPT$ conservation. Consider the ${\cal A}^-=\bra{\lambda}\mathcal{H}_W\ket{h}$ decay amplitudes 
of a meson $h$ into a hadron state $\lambda$ and its $CP$ conjugated process ${\cal A}^+=\bra{\bar \lambda}\mathcal{H}_W\ket{\bar h}$. Here $\mathcal{H}_W$ is the electroweak Hamiltonian. Customarily, we write ${\cal A}^{\pm}=A_\lambda+B_\lambda e^{\pm i\gamma}$,
where $A_\lambda,B_\lambda$ are $CP$ invariant and only the weak phase $\gamma$ sign changes under $CP$. However, when the final state $\lambda$ is coupled to other physically accessible states $\lambda'$, we could consider that it has been produced directly from the source or via another intermediate state. 
Formally, to the lowest order effect due to FSI, we write \cite{Bediaga:2013ela, Cheng:2020ipp}  
\begin{equation}
    {\cal A}^\pm_{LO}=A_\lambda+B_\lambda e^{\pm i\gamma}+i\sum\limits_{\lambda'} \hat f_{\lambda'\lambda} \left(A_{\lambda'}+B_{\lambda'} e^{\pm i\gamma} \right),
    \label{eq:simpleALO}
\end{equation}
where  ${\hat f}_{\lambda\lambda'}$ is the 
 two-body scattering partial wave related to the $S$-matrix  by $S_{\lambda\lambda'}=\delta_{\lambda\lambda'}+2i {\hat f}_{\lambda\lambda'}$. The factor of 2 in front of $\hat f$, absent in \cite{Bediaga:2013ela}, is standard for partial waves and essential \cite{Suzuki:1999uc,Cheng:2020ipp} to arrive to Eq.\eqref{eq:simpleALO}. 
Now $A_\lambda$ and $B_\lambda$ are understood as  decay amplitudes without FSI. The above expression is formal and  $\lambda'$ represents the two particles that re-scatter in a 
definite spin and isospin state.  
$CP$ asymmetries are then defined through $\Delta\Gamma_\lambda=\Gamma_{h\to\lambda}-\Gamma_{\bar h\to\bar\lambda}$ with, generically, $\Gamma=\vert {\cal A}_{LO}\vert^2$.

These processes involve three final mesons, but there is strong evidence that, at least in some regions of phase space, the first two-body scattering largely dominates the FSI \cite{Magalhaes:2011sh} and the other meson acts as a spectator. 
Four or more meson intermediate states are  negligible below CM energies of 1 GeV and relatively small up to roughly 1.5 GeV,  where giant CPV is observed.

Following \cite{Bediaga:2013ela}, let us consider for now just the isoscalar $S$-wave $\pi\pi\leftrightarrow K \bar K$ rescattering, i.e. $\lambda=\pi\pi$, $\lambda'=KK$, which is the most interesting contribution to $\Delta\Gamma_{KK(\pi\pi)}$ in $B^\pm\to K^\pm K^+ K^- (K^\pm \pi^+\pi^-)$, when the two-meson invariant mass is in the 1 to 1.5 GeV range. We will add the other terms and waves later. Since $2\,i{\hat f}_{\pi\pi KK}=S_{\pi\pi KK}=\vert S_{\pi\pi KK}\vert \exp(i\delta_{\pi\pi KK})$,  we can write:
\begin{equation}
\Delta\Gamma_{KK}\simeq {\cal C} \vert S_{\pi\pi KK}\vert \cos(\delta_{\pi\pi KK}+\Phi_{KK})F(M_K^2).
\label{eq:compound}
\end{equation}
 Following \cite{Bediaga:2013ela} we define
${\cal C}=4\vert K\vert \sin\gamma$, where
$K=\vert K\vert \exp(i \Phi_{KK})=B^*_{KK}A_{\pi\pi}-B^*_{\pi\pi}A_{KK}$, with
$K_{KK}=-K_{\pi\pi}$ and $\Phi_{KK}=\Phi_{\pi\pi}+\pi$ due to CPT. Within this first approximation, ${\cal C}$ can be considered constant compared to the strong $s$-dependence of $S_{\pi\pi KK}$. The Dalitz form-factor is $F(M_K^2)=(M_K^2)_{\max}-(M_K^2)_{min}$, which are obtained from kinematics. The amplitude symmetrization in the two
like-sign kaons is neglected as low-mass
regions for each neutral $KK$ pair are very separated in phase space.   CPT implies \cite{Bediaga:2013ela} that these rescattering contributions satisfy  $\Delta\Gamma_{KK}=-\Delta\Gamma_{\pi\pi}$.

\begin{figure}
\includegraphics[scale=0.28]{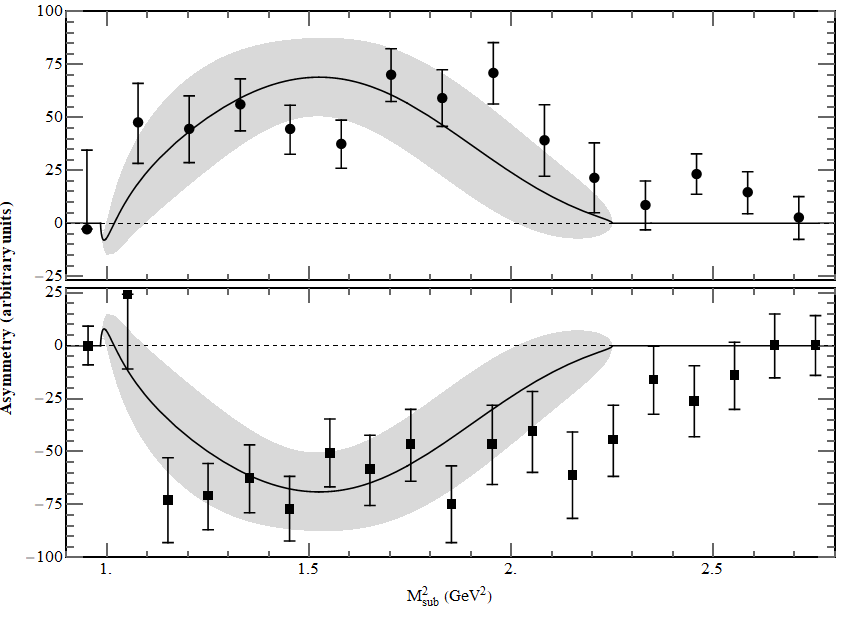}
\caption{\label{fig:plot2014} CP asymmetries for $B^{\pm}\to K^\pm \pi^+\pi^-$ (top) and $B^{\pm}\to K^\pm K^+K^-$ (bottom), from Eq.\eqref{eq:compound} and the estimates
in Eq.\eqref{eq:approx1} and \eqref{eq:approx2}. The plot is identical to Fig.1 in \cite{Bediaga:2013ela}. Data from \cite{LHCb:2013ptu}.}
\end{figure}

However, two crude estimates originally made in \cite{Bediaga:2013ela} have become standard but are not needed and can be easily improved.  Note they have been used in  \cite{Nogueira:2015tsa,Cheng:2016shb,Cheng:2020ipp} and also in the LHCb implementation of this model in \cite{Aaij:2019qps,Aaij:2019hzr,Aaij:2019jaq}. A possible reason for such estimates was that meson-meson scattering are plagued with systematic uncertainties and frequently analyzed with crude models. 
However, at the time of \cite{Bediaga:2013ela} a dispersively constrained analysis existed for $\pi\pi\to\pi\pi$ \cite{Pelaez:2004vs}. Thus, in order to use this dispersive representation, the first approximation was to assume a formalism with only two channels: $1=\pi\pi$ and $2=KK$, so that $S$-matrix  unitarity implies:
\begin{gather}
    \left(S_{\lambda \lambda'}\right)=\begin{pmatrix}
    \eta e^{2i\delta_{11}} &i\sqrt{1-\eta^2}e^{i(\delta_{11}+\delta_{22})} \\
    i\sqrt{1-\eta^2}e^{i(\delta_{11}+\delta_{22})}&\eta e^{2i\delta_{22}}
    \end{pmatrix},
    \nonumber
\end{gather}
where $\eta$ is the $\pi\pi\to \pi\pi$ elasticity. Hence, the required $S_{\pi\pi KK}$ was avoided by replacing in Eq.\eqref{eq:compound}:
\begin{eqnarray}
&&\vert S_{\pi\pi KK}\vert \longrightarrow\sqrt{1-\eta^2},
\label{eq:approx1}\\
&&\delta_{\pi\pi KK}\longrightarrow \delta_{\pi\pi\pi\pi}+\delta_{KK KK}\simeq 2 \delta_{\pi\pi\pi\pi},
\label{eq:approx2}
\end{eqnarray}
where in the last step $\delta_{KKKK}\simeq\delta_{\pi\pi\pi\pi}$, was assumed, since little is known about $\delta_{KKKK}$. Finally, setting $\Phi_{KK}=0$ as well as $\delta_{\pi\pi KK}=0$ above 1.5 GeV, we reproduce in Fig.\ref{fig:plot2014} the results of \cite{Bediaga:2013ela} for $\Delta\Gamma_{KK}$ (bottom) and $\Delta\Gamma_{\pi\pi}$ (top), projected from LHCb results \cite{Aaij:2013sfa}, as a function of the two-meson  invariant mass $M^2_{sub}=s$. Only the normalization constant $\cal C$ is free. Note that as nicely shown in \cite{Bediaga:2013ela}, due to CPT symmetry, and just by changing its global sign,  Eq.\eqref{eq:compound} roughly  describes
both asymmetries from $K\bar K$ threshold  to $M_{sub}\simeq 1.5\,$GeV, i.e., $S$-wave FSI dominate the $s$-dependence in that region.

\begin{figure}
\includegraphics[scale=0.62]{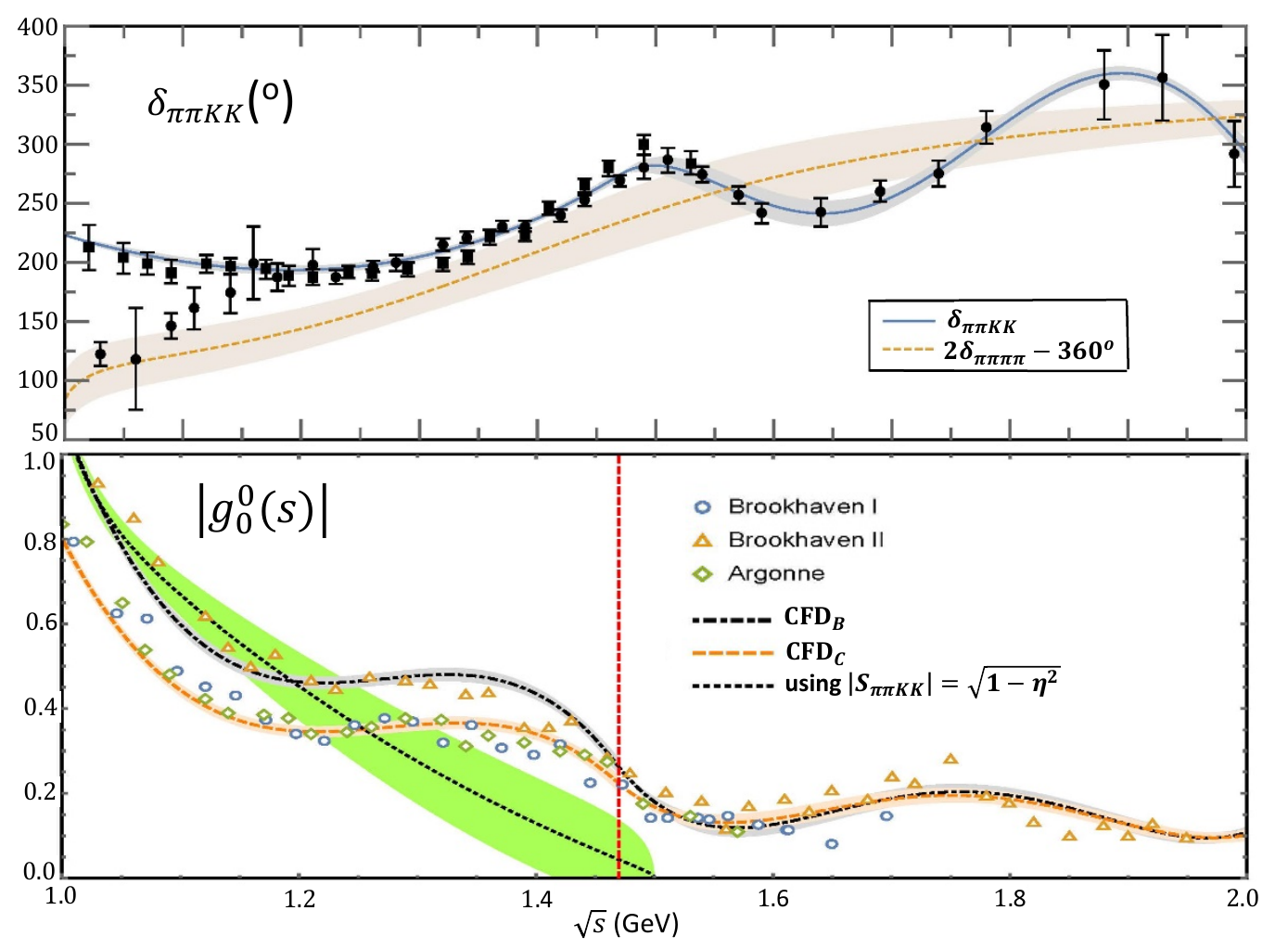}
\caption{\label{fig:PhasePiK} Top: $\delta_{\pi\pi K \bar K}$ data from \cite{Cohen:1980cq} (squares) and \cite{Etkin:1981sg} (circles). The dashed line is the  Eq.\eqref{eq:approx2} estimate, although  subtracting $2\pi$  to fit in the plot, and using \cite{Pelaez:2004vs} (PY) for $\delta_{\pi\pi\pi\pi}$. 
The continuous line is the dispersively constrained fit from \cite{Pelaez:2018qny} (PR). The 5 first data points of \cite{Etkin:1981sg} below $1.2$ GeV are in conflict with Watson's Theorem and dispersive analyses of $\pi\pi\to \pi\pi$ and are commonly discarded.
Bottom:  $\vert g_0^0(s)\vert$ data. The green band is Eq.\eqref{eq:approx1} and the grey and orange bands correspond to the 
dispersive analysis in \cite{Pelaez:2018qny}.}
\end{figure}

However, neither of these estimates is needed because both $\delta_{\pi\pi KK}$ and $\vert S_{\pi\pi KK}\vert$ data up to 2 GeV exist since the 80's from the Argonne \cite{Cohen:1980cq}, Brookhaven-I \cite{Etkin:1981sg} and II \cite{Longacre:1986fh} collaborations, shown in Fig.\ref{fig:PhasePiK}. Note that to compare with data we employ the usual normalization
\begin{equation}
\vert g^0_0(s)\vert=\frac{\sqrt{s}}{4(q_\pi q_K)^{1/2}} \vert S_{\pi\pi KK}(s)\vert,\quad s>4 m_K^2,
\label{eq:modStog}
\end{equation}
with  $q_P=\sqrt{s/4- m_P^2}$ 
the $P=\pi, K$ CM momenta.
There we see that Eqs.\eqref{eq:approx1} and \eqref{eq:approx2}  fail to describe both the
$\vert S_{\pi\pi KK}\vert$ and $\delta_{\pi\pi KK}$ data, respectively. 
For these curves we use the $\delta_{\pi\pi \pi\pi}$ 
and $\eta$ obtained in \cite{Pelaez:2004vs}, because it has become customary in the literature, although those dispersively constrained fits were updated in \cite{Garcia-Martin:2011iqs}. Had we used the latter, 
with smaller uncertainties,
the comparison with data would be even worse. Recall also that we are  subtracting $2\pi$ \cite{footnote1}
to make $2\delta_{\pi\pi\pi\pi}$ fit in the plot.
Hence,  Eqs.\eqref{eq:approx1} and \eqref{eq:approx2}
should be avoided.
But then one might wonder if the claimed relevance of $\pi\pi\to K \bar K$ FSI in giant CPV depends crucially on such crude estimates and their large uncertainties, or if they still hold when a realistic $\pi\pi\to K \bar K$ parameterization is used instead.

Luckily, only very recently, but very timely, model-independent
dispersive analyses of $\pi\pi\to K K$ data, using hyperbolic dispersion relations (Roy-Steiner equations), have become available in \cite{Pelaez:2018qny} and updated in \cite{Pelaez:2020gnd}. These provide accurate Constrained Fits to Data (CFD) up to 1.47 GeV, the maximum applicability of these relations, continuously matched to unconstrained fits up to 2 GeV, for both $\delta_{\pi\pi KK}$ and $\vert S_{\pi\pi KK}\vert$, shown in  Fig.\ref{fig:PhasePiK}. Note that for the modulus there are two solutions. We will present results for the higher one since their 
difference up to 1.47 GeV can be reabsorbed in the normalization parameter and at the end yield very similar results.

\begin{figure}
\includegraphics[scale=0.28]{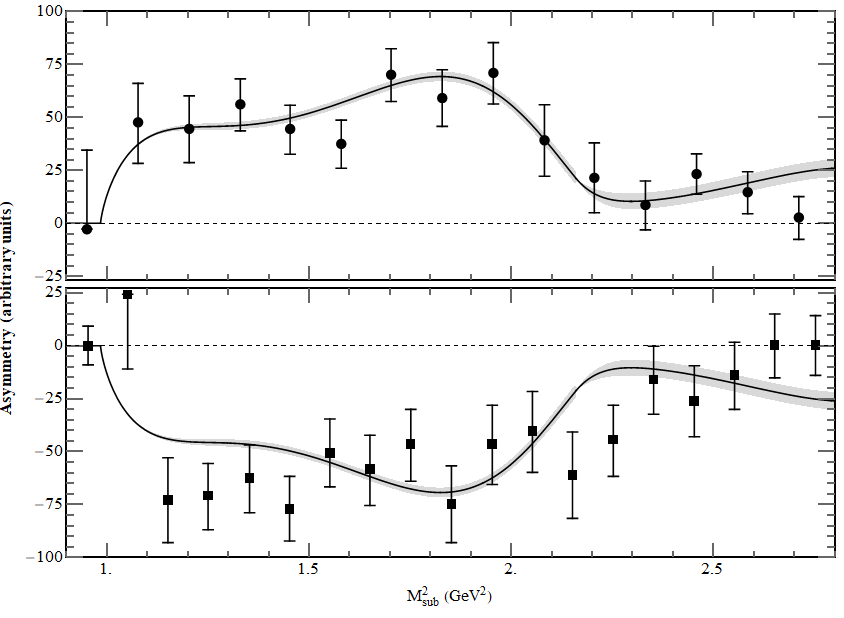}
\caption{\label{fig:plot2014new}
As Fig.\ref{fig:plot2014} but using in Eq.\eqref{eq:compound} the dispersively constrained CFD parameterization of $\pi\pi\to K\bar K$ data from \cite{Pelaez:2018qny,Pelaez:2020gnd}. Note the huge increase in precision with respect to Fig.\ref{fig:plot2014} and the new patterns due to resonance interplay.
 }
\end{figure}
Thus, in Fig.\ref{fig:plot2014new}
we show the asymmetry results when the CFD $\pi\pi\to K \bar K$ dispersive analysis is
used in Eq.\eqref{eq:compound}. 
Our $\chi^2_{\rm d.o.f}=1$ and we have freed the parameter $\Phi$ finding a nonvanishing preferred value $(-34\pm9)^{\rm o}$. There is an impressive improvement in precision with respect to Fig.\ref{fig:plot2014}.
Remarkably, it also shows peaks and dips associated to the interplay of the $f_0(980)$, $f_0(1370)$ and $f_0(1500)$ resonances \cite{Pelaez:2022qby} that were concealed in Fig.\ref{fig:plot2014} within the large uncertainty  from Eqs.\eqref{eq:approx1} and \eqref{eq:approx2}. 

\begin{figure}
\includegraphics[scale=0.28]{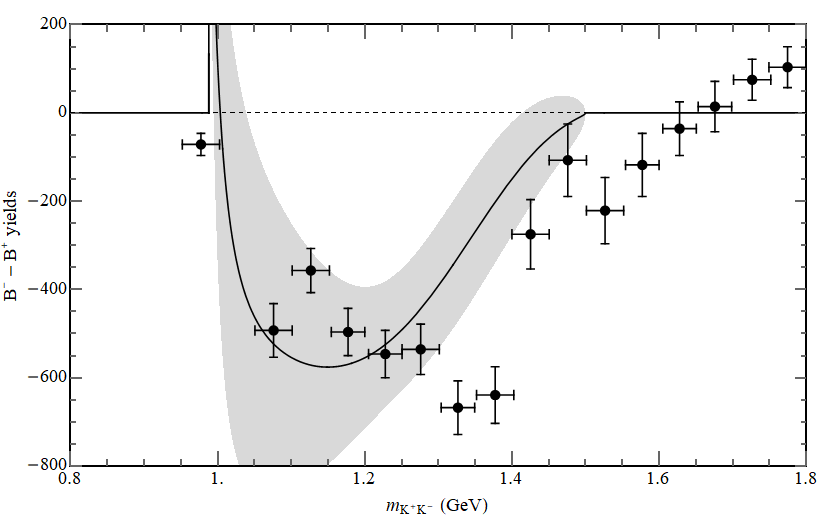}
\caption{\label{fig:plot2015} Total  $B^{\pm}\to K^\pm K^+K^-$ asymmetry.
LHCb data from the sum of Figs 6(c) and (d) in \cite{Aaij:2014iva}. Central line, using Eqs.\eqref{eq:approx1} and \eqref{eq:approx2}, identical to  \cite{Nogueira:2015tsa}. We have added here the huge uncertainty in that description. }
\end{figure}

\begin{figure}
\includegraphics[scale=0.28]{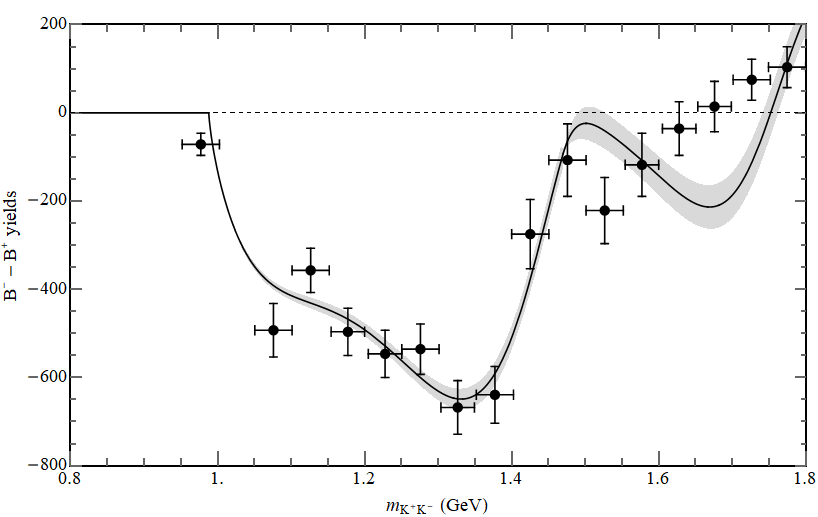}
\caption{\label{fig:plot2015new} As in Fig.\ref{fig:plot2015} but using the dispersively constrained fit to $S_{\pi\pi KK}$ data in \cite{Pelaez:2018qny}.
 Note the dramatic improvement in precision and the unveiling of resonant structures.}
\end{figure}

Furthermore, the full run I LHCb data on the $B^\pm\to K^\pm K^+K^-$ CPV asymmetry 
\cite{Aaij:2014iva}, were described  in \cite{Nogueira:2015tsa} with Eq.\eqref{eq:compound} above, but divided by $(1+s/\Lambda_\lambda^2)(1+s/\Lambda_{\lambda'}^2)$, to mimic the mild $s$-dependence of the source term for each $\lambda$ pair, with $\Lambda_{KK}=4\,$GeV and $\Lambda_{\pi\pi}=3\,$GeV. 
Using then Eqs.\eqref{eq:approx1} and \eqref{eq:approx2}, we have reproduced in Fig.\ref{fig:plot2015} the central value of  \cite{Nogueira:2015tsa}, but also adding the huge uncertainty due to such estimates.
In contrast, we show in Fig.\ref{fig:plot2015new} the result when using the CFD phase and modulus in \cite{Pelaez:2018qny,Pelaez:2022qby}. The central line
 follows much better the data, with dramatically smaller uncertainties, again unraveling the interplay between  resonances. See details in
 the Appendix A. Above 1.5 GeV this approach is not expected to be valid, due to the increasing relevance of $4\pi$ and  other resonances and FSI with higher angular momenta.

All in all, these results confirm, using realistic and accurate FSI, 
that $\pi\pi\to K \bar K$ rescattering does indeed play a dominant role in the appearance and the $s$-dependence of giant CPV at LHCb in the 1 to 1.5 GeV region.

\begin{figure}
\hspace*{-3mm}
\includegraphics[scale=0.447]{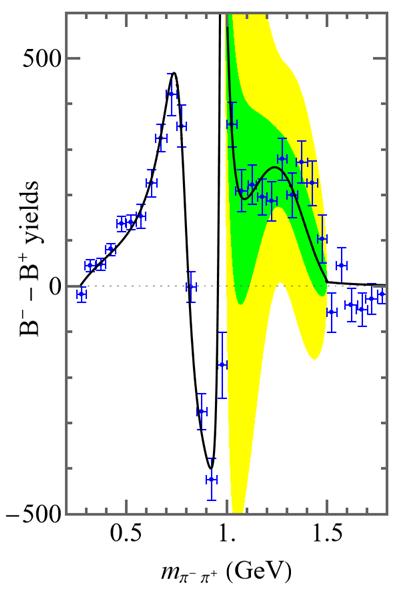}
\includegraphics[scale=0.447]
{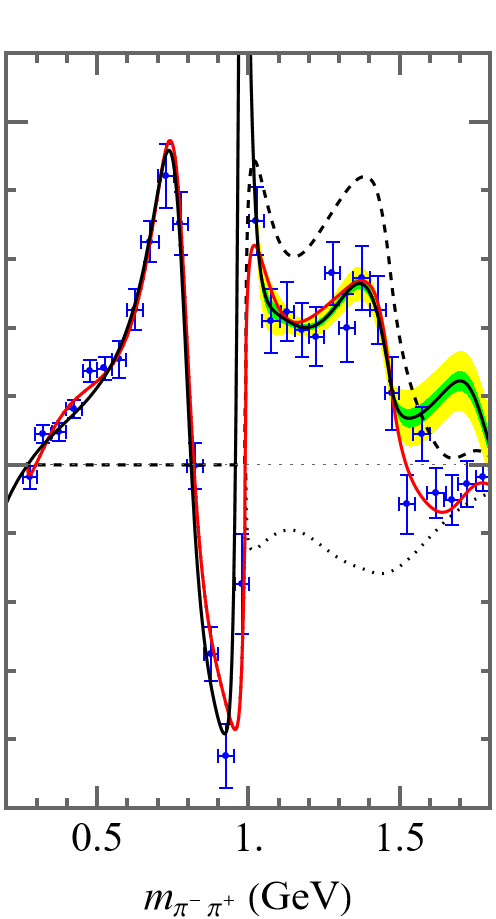}
\caption{\label{fig:plotKpipi}  $B^{\pm}\to K^\pm \pi^+\pi^-$ asymmetry in the $\cos{(\theta)}<0$ region. Left: the central value reproduces Fig.11 in \cite{Nogueira:2015tsa}. We have added here the one (green) and three (yellow) standard deviation bands due to the crude estimates in Eq.\eqref{eq:approx1} and \eqref{eq:approx2}. Right: Same but using the dispersive $\pi\pi\to K\bar K$ data analysis in \cite{Pelaez:2018qny}  (black line), as proposed here. The accuracy improvement is dramatic.  In addition, we show 
in red the full $\Delta \Gamma$ (Eq.44 in \cite{Nogueira:2015tsa}) but with the dispersive $\pi\pi \to \pi\pi$ isoscalar S-wave in \cite{Pelaez:2019eqa} instead of just a  $f_0(980)$ Breit-Wigner. The high-mass region is now well described and the too-large $f_0(980)$ peak disappears, while including the $f_0(500)$ and $f_0(1370)$. In the 1-1.5 GeV region the terms containing the $\pi\pi \to KK$ amplitude (dashed)
largely dominate those without it (dotted).}    
\end{figure}

Let us now reintroduce other relevant terms, following the more complete model of \cite{Nogueira:2015tsa}, adopted by the LHCb analyses  \cite{Aaij:2019qps,Aaij:2019hzr,Aaij:2019jaq}. Thus, we recast
Eq.\eqref{eq:simpleALO} as:
\begin{eqnarray}
{\cal A}^\pm_{LO}
&=&
\sum\limits_{J} 
(a_{\lambda NR}^{J}+b_{\lambda NR}^J e^{\pm i\gamma})/(1+s/\Lambda_\lambda^2) \label{eq:ALO}\\
&+&
\sum\limits_{JR} 
(a_\lambda^{R}+b_\lambda^{R} e^{\pm i\gamma})
F^{BW}_{R\lambda}P_J(\cos \theta)\nonumber\\
&+&i\sum\limits_{\lambda',J}
 {\hat f}_{\lambda'\lambda}^J \left(a_{\lambda' NR}^J+b_{\lambda' NR}^J e^{\pm i\gamma} \right)/(1+s/\Lambda_{\lambda'}^2),
\nonumber
\end{eqnarray}
where the angular momenta $J$ is explicitly separated from $\lambda$. 
Note that terms without FSI and
other mild $s$-dependent contributions are grouped into a non-resonant (NR) part.
Besides, the strong $s$-dependence of elastic scattering, $\lambda'=\lambda$, is described with usual Breit-Wigner shapes. Namely, $(1+i\hat f^J_{\lambda\lambda})A^J_{\lambda R}\to a_0^R F^{BW}_R P_J(\cos\theta)$, with $\theta$ the helicity angle between the like-sign
mesons in the Gottfried-Jackson frame, 
and 
\begin{gather}
    F^{\textnormal{BW}}_R=\frac{1}{m_R^2-s-im_R\Gamma_R(s)}, 
   \,\Gamma_R(s)=\frac{q_\pi(s)m_R\Gamma_R}{q_\pi(m_R^2)s^{1/2}}.
    \nonumber
\end{gather}
The two first terms in Eq.\eqref{eq:ALO} correspond to a familiar isobar model, whereas inelastic FSI appear in the third term, dominated by  $\hat f^0_{\pi\pi KK}$. The resonances to be considered depend on the process.
For instance, for the $B^\pm \to K^\pm\pi^+\pi^-$ asymmetries, the energies below the $KK$ threshold become accessible. In \cite{Nogueira:2015tsa}
the $J=1,0$ waves were 
 approximated only with the $\rho(770)$ and $f_0(980)$ resonances, respectively, by setting
 \begin{equation}
    A_{\lambda R}=a_{\lambda}^\rho F_\rho^{\textnormal{BW}}(s)k(s)\cos{(\theta)}+a^f_\lambda F_f^{\textnormal{BW}}(s),
    \label{eq:rhoandf}
\end{equation}
 with $k(s)= \sqrt{1-4m_\pi^2/s}$  and similarly for $B_{\lambda R}$ amplitudes.  Thus, on the left panel of Fig.\ref{fig:plotKpipi} we reproduce
 the central value of the partial CPV asymmetry obtained in \cite{Nogueira:2015tsa} with this improved model, and using Eqs.\eqref{eq:approx1} and \eqref{eq:approx2}. Note the nice $\rho(770)$ resonant peak and dip structure around 770 MeV and the marked peak of the $f_0(980)$. However, once again we are providing the huge uncertainties that appear due to such standard crude estimates.
 In contrast, on the right panel, we show the remarkable accuracy attained when using the dispersive $\pi\pi\to K\bar K$ amplitude instead. Details can be found in Appendix A.

 So far we have limited ourselves to the crude
but appealing model formulated in \cite{Nogueira:2015tsa}, in order to show the accuracy  improvement when using
the recent dispersively constrained parameterizations of $\pi\pi\to K \bar K$
in the same kind of analyses that had been widely used before.
It is enough to restore back all the instances of $2\delta_{\pi\pi\pi\pi}$ and $\sqrt{1-\eta^2}$ by, respectively, the $\delta_{\pi\pi KK}$  and $\vert S_{\pi\pi KK} \vert$ parameterizations \cite{footnote2}
in \cite{Pelaez:2018qny,Pelaez:2020gnd}.

However, the accuracy attained in $S$-wave FSI, 
opens an interesting outlook for further studies. Hence, reconsidering contributions neglected so far becomes even more appealing. 
The proponents of this model already pointed out some possible improvements, particularly  the inclusion of a  realistic $\pi\pi$ $S$-wave.
Thus Fig.\ref{fig:plotKpipi} shows in red  the result of replacing the naive single $f_0(980)$  Breit-Wigner shape with the dispersive $\pi\pi$ data analysis in \cite{Pelaez:2019eqa}, which also describes the $f_0(500)$ and $f_0(1370)$.
The contributions containing the $\pi\pi\to K\bar K$ amplitudes still dominate the 1 to 1.5 GeV region.
Details are provided in Appendix B.
Further waves and resonances could also be implemented in a similar way. 

Our level of precision calls for a future replacement of this simple isobar model with leading order FSI corrections 
by the full treatment of the three-body decay, but containing the correct two-body rescattering amplitude. 
A first step would be to consider the all-order two-body contributions \cite{Suzuki:1999uc,Cheng:2016shb,Cheng:2020ipp}, although eventually it should include dispersively constrained third-particle effects following what has been done for $D$ or other heavy meson decays \cite{Niecknig:2015ija,Niecknig:2017ylb, Akdag:2021efj}, but restricted to certain regions of the $B$-decay phase space. Finally, three-body contributions should be included. For ongoing efforts in these topics we refer to \cite{Magalhaes:2011sh, Nakamura:2015qga,JPAC:2019ufm} and references therein. 
This work, therefore, paves the way for several future developments.

In summary, we have shown how to implement the recent dispersively constrained parameterizations of $\pi\pi\to K\bar K$ to describe final state interactions in charmless three-body B decays, avoiding standard crude estimates within a popular model used by LHCb
and others to describe giant CP violation. This will help reduce by about an order of magnitude the uncertainty due to FSI in that
model. As a result, the dominant role of inelastic final state interactions in the strong $s$-dependence measured in the 1 to 1.5 GeV 
region, previously based on crude estimates, is confirmed with realistic interactions. 
Moreover, this dramatic reduction of uncertainty when using the dispersive analyses of $\pi\pi\to K\bar K$ data, opens the way for a more detailed description of additional hadronic features and a precise treatment of further data. This is particularly relevant
in the amplitude analysis that should be carried out for data just released by LHCb \cite{LHCb:2022nyw} or to be obtained in the near future.
\section*{Acknowledgements}
We thank I. Bediaga, T. Frederico and 
H.-Y. Cheng, for discussions and clarifications. 
Research partially funded by the Spanish Ministerio de Ciencia e Innovaci\'on grant PID2019-
106080GB-C21 and the European Union’s Horizon 2020
research and innovation program under grant agreement
No 824093 (STRONG2020).  PCM was supported by Spanish Ministerio de Ciencia e Innovaci\'on grant Maria Zambrano para atracción de talento interancional (Convocatoria 2021-2023).

\appendix
\section{Details of Calculations}
\label{Appendix}

To illustrate the accuracy achieved by using the dispersive analysis of $\pi\pi\to K\bar K$ data in \cite{Pelaez:2018qny,Pelaez:2020gnd} instead of the standard crude estimates, we have made use in our calculations  of the very same model employed by the LHCb own analyses of giant CPV \cite{Aaij:2019qps,Aaij:2019hzr,Aaij:2019jaq}, which was introduced by \cite{Nogueira:2015tsa} (note that for the specific case of $B\to \pi\pi\pi$, other models have been used in \cite{Aaij:2019hzr,Aaij:2019jaq}). In particular, after separating the amplitude in non-resonant and resonant parts, Eq.\eqref{eq:ALO}, and considering just the $\rho(770)$ and $f_0(980)$ resonances as in Eq.\eqref{eq:rhoandf}, the amplitude of the process is given by $\Gamma=\vert {\cal A}_{LO}\vert ^2$. Then, $CP$ asymmetries are defined as $\Delta\Gamma_\lambda=\Gamma_{h\to\lambda}-\Gamma_{\bar h\to\bar\lambda}$. With a little bit of algebra, all the dependence in the generic $a_\lambda,b_\lambda$ coefficients can be recast in terms of a few constants  multiplying simple functions carrying the $s$-dependence. 
This is nicely done in \cite{Nogueira:2015tsa} and will not be repeated here.
Just for illustration, in the case of $B^{\pm}\to K^\pm K^+K^-$ and $B^{\pm}\to K^\pm \pi^+\pi^-$ decays, 
the asymmetry was reduced to
\begin{widetext}
\begin{align}
\Delta\Gamma_\lambda=&\frac{\mathcal{A}}{\left(1+\frac{s}{\Lambda_\lambda}\right)^2}+
    \frac{\left(\mathcal{B}\cos{(\delta_{\pi\pi KK})}
    +\mathcal{B}'\sin{(\delta_{\pi\pi KK})}
    \right)\vert S_{\pi\pi KK}\vert}{\left(1+\frac{s}{\Lambda^2_\lambda}\right)\left(1+\frac{s}{\Lambda^2_{\lambda'}}\right)}
\nonumber\\
    &
    +\left|F_\rho^{\textnormal{BW}}(s)\right|^2k(s)\cos{(\theta)}\left[\frac{\mathcal{D}(m_\rho^2-s)}{\left(1+\frac{s}{\Lambda^2_\lambda}\right)}+\frac{\mathcal{D}'\vert S_{\pi\pi KK}\vert\left(m_\rho \Gamma_\rho(s)\cos{(\delta_{\pi\pi KK})-(m_\rho^2-s)\sin{(\delta_{\pi\pi KK})}}\right)}{\left(1+\frac{s}{\Lambda^2_{\lambda'}}\right)}\right] 
    \nonumber \\
    &
    +\left|F_\rho^{\textnormal{BW}}(s)\right|^2\left|F_f^{\textnormal{BW}}(s)\right|^2k(s)\cos{(\theta)}\left(\mathcal{F}\left((m_\rho^2-s)(m_f^2-s)+m_\rho\Gamma_\rho(s) m_f\Gamma_f(s)\right)\right.
    \nonumber\\
    &
    \left.+\mathcal{G}\left((m_\rho^2-s)m_f\Gamma_f(s)-m_\rho\Gamma_\rho(s) (m_f^2-s)\right)\right),
\label{eq:fullmodel}
\end{align}
\end{widetext}
with $\lambda' \neq \lambda$. In the model of \cite{Nogueira:2015tsa}, there are more terms  of relevance for other processes or non-invariant under CPT, but in this section we will only keep the very same ones used in that reference, i.e. ${\cal A, B, B', D, D', F, G}$, to ease the comparison, 
particularly on the uncertainties.
The only exception is   ${\cal  B'}$, that was also set to zero but will be
 traded for a $\Phi$ angle below, to ease the comparison with the simplified model in \cite{Bediaga:2013ela},
which corresponds to Eq.\eqref{eq:compound} 
in the main text.
In the next section, we will consider a more complete and improved version of this model.

Note that, contrary to \cite{Nogueira:2015tsa} and the standard use in the literature, in Eq.\eqref{eq:ALO}, we have not made the replacements
\begin{eqnarray}
&&\vert S_{\pi\pi KK}\vert \longrightarrow\sqrt{1-\eta^2},\\
&&\delta_{\pi\pi KK}\longrightarrow  2 \delta_{\pi\pi\pi\pi},
\end{eqnarray}
where $\delta_{\pi\pi\pi\pi}$  and $\eta$ are the phase and elasticity of $\pi\pi\to \pi\pi$ scattering.
Instead, as explained in the main text, one has to use directly the dispersively constrained fits to data (CFD) parameterizations for the $S$-wave isoscalar partial wave $\vert S_{\pi\pi KK}\vert$ and $\delta_{\pi\pi KK}$ obtained in \cite{Pelaez:2018qny} and slightly updated in \cite{Pelaez:2020gnd}. 
Beware, however, of the different notation. In particular, what we call here $\delta_{\pi\pi KK}(s)$ is called $\Phi^0_0(t)$ in \cite{Pelaez:2018qny,Pelaez:2020gnd}. This is because $\pi\pi\to K\bar K$ amplitudes can be obtained from $\pi K\to \pi K$ amplitudes by $s\leftrightarrow t$ crossing. For our purposes in this work, just replace $t$ in \cite{Pelaez:2018qny,Pelaez:2020gnd} by our $s$ (the two-meson subsystem invariant mass) here.

For the modulus, we have to take into account that in  \cite{Pelaez:2018qny,Pelaez:2020gnd} the $S$-wave isoscalar partial-wave amplitude, called $g^0_0(s)$ instead of $\hat f^0(s)$ as in the main text here, was defined with the following normalization:
\begin{equation}
S_{\pi\pi KK}(s)=S^0_0(s)=i\frac{4(q_\pi q_K)^{1/2}}{\sqrt{s}} g^0_0(s),\quad s>4 m_K^2,
\end{equation}
where  $q_P=\sqrt{s/4- m_P^2}$  
are the $P=\pi, K$ CM momenta.
The detailed expression and parameters for $\vert g^0_0(s)\vert$ can be found
in \cite{Pelaez:2018qny,Pelaez:2020gnd}, just replace $t$ there by $s$ here. 

Concerning the parameters of the CFD parameterization, for concreteness, we have used here the CFD$_B$ parameterization, but we have also tried CFD$_C$. Note that their phases are almost identical and they only differ in the modulus, which is somewhat smaller for CFD$_C$, but has the same peaks and dips structure. However, that is easily absorbed in the parameters of our fit here, particularly in those affecting normalizations. 
Recall that, intuitively, the FSI dominates the $s$-dependence.
As a consequence,
the results for the fits of the asymmetries, using either CFD$_B$ or CFD$_C$, are almost indistinguishable to the eye, which is why we have only shown results for CFD$_B$. 
Finally, let us comment that the values of the CFD$_B$ and CFD$_C$ parameters in \cite{Pelaez:2018qny} have been upgraded in \cite{Pelaez:2020gnd} by imposing dispersion relations not only for $\pi\pi\to K \bar K$ but also for the simultaneous fits to the cross-channel $\pi K\to \pi K$ process. The difference between the values of the parameters in \cite{Pelaez:2018qny} and \cite{Pelaez:2020gnd} is in general minute for the CFD parameterizations. 
Note that everything concerning rescattering and FSI is fixed in the fits, i.e. pure input. Only the relevant ${\cal A, B, B', C, D, D', F, G}$ for each process are fitted.

Finally, in this work, we have calculated asymmetries
either 
partially or totally integrated with respect to $\cos \theta$. As nicely derived in \cite{Nogueira:2015tsa}, the three body kinematics of three particles labeled 1,2,3, 
implies that this cosine can be recast as:
\begin{gather}
    \cos{(\theta)}=a(s)m_{23}^2+b(s),\nonumber \\
    a(s)=\frac{1}{\left(s-4m_\lambda^2\right)^{1/2}\left(\frac{(m_B^2-m_K^2-s)^2}{4s}-m_K^2\right)^{1/2}}, \nonumber \\
    b(s)=\frac{a(s)}{2}\left(s-m_B^2-m_K^2-2m_\lambda^2\right).
\end{gather}
with $ m_B=5.279\,$GeV, $s=m^2_{12}$ the invariant mass squared of the 1-2 subsystem (the one that re-scatters), $m^2_{23}$ the one of the 2-3 subsystem and $m_\lambda=m_K$ or $m_\pi$ depending on what is the bachelor particle. Hence, the  asymmetry now depends on two variables $\Delta\Gamma_\lambda(s,m_{23}^2)$,
leading to the total asymmetry $\Delta\Gamma_\lambda(s)=\int \Delta\Gamma_\lambda(s,m_{23}^2)dm_{23}^2$.

None of these complications exist for the simplest model of Eq.\ref{eq:compound}, used for Figs.\ref{fig:plot2014} and \ref{fig:plot2014new}, since there we just concentrated in the scalar case. The same happens in the calculation of the $B^\pm\to K^\pm K^+K^-$ asymmetry and Figs.\ref{fig:plot2015}  and \ref{fig:plot2015new}. Since we wanted to illustrate the dramatic improvement due to avoiding crude estimates, we have followed the approach of  \cite{Nogueira:2015tsa} and considered only the ${\cal B, B'}$ terms.
As a matter of fact, in \cite{Nogueira:2015tsa} ${\cal B'}$ is also set to zero, although as commented above and in \cite{Nogueira:2015tsa}, ${\cal B}$ and ${\cal B'}$ can be traded for just one   normalization constant, say ${\cal \hat B}$, and a constant angle $\Phi$, using
\begin{gather}
\mathcal{\hat{B}}\cos{(\delta_{\pi\pi KK}+\Phi)}=\mathcal{B}\cos{(\delta_{\pi\pi KK})}+\mathcal{B}'\sin{(\delta_{\pi\pi KK})}.
    \label{eq:cosrelation}
\end{gather}
The above  equation is just a constant shift on the phase, which  does not change the overall size of the uncertainty bands.
Note also that, in the $\Lambda_{\pi\pi}, \Lambda_{KK}\to\infty$ limits, this reduces to Eq.\eqref{eq:compound}. Namely, we are only considering the terms with $S$-wave FSI. Given the small uncertainties of the dispersively constrained CFD parameterization of $\pi\pi\to KK$, the data description with this simple model is still impressive.
Moreover, the $\chi^2_{\textnormal{d.o.f.}}$ calculated with
the central values of the CFD up to 1.5 GeV
is 1.9 whereas it was 8.7 if using the crude estimate as in Fig.\ref{fig:plot2015}.
The resulting parameters   for the fit in Fig.\ref{fig:plot2015new}  are ${\cal \hat B}=62\pm7$ and $\Phi=(-25\pm19)^{\rm o}$.
This large improvement in  $\chi^2_{\textnormal{d.o.f.}}$ would not be possible without the pattern of resonance interplay present in the $\pi\pi\to K\bar K $
dispersive analysis, which allows for describing
the data points much better between 1.3 to 1.5 GeV while lying relatively close to those around 1.1 GeV. With this level of accuracy, one could now think about adding further structure and subdominant contributions to the model.

This is why we have considered all the other terms in Eq.\eqref{eq:fullmodel} to describe
the $B^\pm\to K^\pm\pi^+\pi^-$ asymmetry measured in \cite{Aaij:2014iva}.
In this case, as already discussed in \cite{Nogueira:2015tsa}, we only consider the partial asymmetry for $\vert \cos \theta\vert<1$, since in the other region there is a possible presence of
rescattering coming from double charm decays, absent in this model \cite{TIP_kkk,TIP_ppp}.  Thus, we have to describe the partially integrated asymmetry defined as 
\begin{equation}
\Delta\Gamma(s)^{(\cos\theta<0)}=\int_{-\frac{1+b}{a}}^{-\frac{b}{a}} 
 \Delta\Gamma(s,m_{23}^2)
dm^2_{23},
\end{equation}
where, for clarity, in the integration limits we have suppressed the $s$-dependence of $a(s)$ and $b(s)$.
Once again, we refer to \cite{Nogueira:2015tsa} for the detailed derivation. 

These are the equations we have used to obtain Fig.\ref{fig:plotKpipi}.
However, we should note that we cannot reproduce an additional $1/\sqrt{s-4m_P^2}$ threshold factor that
is present in the equations of \cite{Nogueira:2015tsa} after integration.
Thus our calculations do not have it. 

The absence of that factor can be illustrated
by integrating the simplest term in Eq.\eqref{eq:fullmodel} for the $KKK$ asymmetry. Namely,
${\cal A}/(1+s/\Lambda_\lambda^2)^2$,
 which does not depend on $\cos\theta$ and therefore on $m_{23}$. Thus, it factors out of the integral
\begin{eqnarray}
\Delta\Gamma(s)^{(\cos\theta<0)}&=&\int_{-\frac{1+b}{a}}^{-\frac{b}{a}} 
 \frac{\mathcal A}{\left(1+\frac{s}
     {\Lambda_\lambda^2}\right)^2}
dm^2_{23}\nonumber\\
&=& \frac{\mathcal A}{\left(1+\frac{s}
     {\Lambda_\lambda^2}\right)^2}\int_{-\frac{1+b}{a}}^{-\frac{b}{a}} dm^2_{23}
=\frac{{\mathcal A}/a(s)}{\left(1+\frac{s}
     {\Lambda_\lambda^2}\right)^2},
     \nonumber
\end{eqnarray} 
Whereas in \cite{Nogueira:2015tsa} it is used
  \begin{equation}
     \frac{{\mathcal A}/a(s)}{\sqrt{s-4m_K^2}\left(1+\frac{s}
     {\Lambda_\lambda^2}\right)^2}.
 \end{equation}
 As commented above we have not been able to find that extra $\sqrt{s-4m_K^2}$, which thus is not present in our calculations. 
 
 For completeness, we provide here the values of the parameters of our fit, up to 1.5 GeV, for the $B^\pm\to K^\pm\pi^+\pi^-$ and the right panel of Fig.\ref{fig:plotKpipi}: 
\begin{eqnarray}
 {\cal \hat B}=-72\pm6,&\quad& \Phi=(-66\pm13)^{\rm o},\nonumber\\
 {\cal D}=-1.8\pm0.9, &\quad&{\cal D'}=-91\pm10\nonumber\\
 {\cal F}=-4.5\pm0.8, &\quad& {\cal G}=1.7\pm0.4,  \nonumber
 \end{eqnarray}
 Once again we have traded ${\cal B}$, and ${\cal B'}$ for ${\cal \hat B}$ and $\Phi$, respectively, using Eq.\eqref{eq:cosrelation}.   
As seen in Fig.\ref{fig:plotKpipi}, the fit clearly shows the observed structures and is fairly good up to 1.5 GeV, where other contributions are expected to become relevant.
The $\chi^2_{\textnormal{d.o.f.}}=1.5$, so, there is still room for improving the model with
further resonances, like the $f_0(500)$ at low energies, or the $\rho'(1450)$, as well as a FSI treatment of higher partial waves similar to that performed here for the $\pi\pi\to K\bar K$ $S$-wave.
In particular, the proponents of the model \cite{Bediaga:2013ela}, already suggested several modifications of which the first was an improved treatment of the $\pi\pi$ $S$-wave,
which we will discuss next, although it has not been included in the LHCb implementation of this model.

\section{Realistic $\pi\pi\to \pi\pi$ scalar isoscalar wave}
\label{AppendixB}

The first suggested improvement of the previous model is to replace the single $f_0(980)$ Breit-Wigner 
description of the  $\pi\pi$ scattering isoscalar $S$-wave by 
the ``global" parameterization of \cite{Pelaez:2019eqa}. This amplitude describes the existing $\pi\pi$ data while satisfying partial-wave Roy-like dispersion relations up to 1.1 GeV as well as Forward dispersion relations up to 1.42, while containing poles for the $f_0(500)$, or $\sigma$,  $f_0(980)$ and $f_0(1370)$ scalar resonances. From 1.4 to 2 GeV
it is just a fit to data. 
We showed the result of this replacement  as a red curve in the right panel of Fig.\ref{fig:plotKpipi}. Note the disappearance of the artificially large
$f_0(980)$ peak
and the remarkable improvement in the region above 1.5 GeV.

In addition, now we also consider the full interference pattern between $NR$, $\rho$, and  $S$-wave amplitudes (Eq.(44) of \cite{Nogueira:2015tsa}), besides the terms
in Eq.(\ref{eq:fullmodel}). All in all, the improved model reads: 

\begin{widetext}
\begin{eqnarray}    
\Delta\Gamma&=&
\frac{\mathcal{A}}{\left(1+\frac{s}{\Lambda_\lambda}\right)^2}
+\frac{\mathcal{\hat B}\cos{(\delta_{\pi\pi KK}+ \Phi} )\vert S_{\pi\pi KK}\vert}{\left(1+\frac{s}{\Lambda^2_\lambda}\right)\left(1+\frac{s}{\Lambda^2_{\lambda'}}\right)}
+\mathcal{C}\left|F_\rho^{\textnormal{BW}}(s)\right|^2k^2(s)\cos^2{(\theta)}
\nonumber\\    
    &+&\left|F_\rho^{\textnormal{BW}}(s)\right|^2k(s)\cos{(\theta)}
    \Bigg[
    \frac{\mathcal{D}(m_\rho^2-s)}{\left(1+\frac{s}{\Lambda^2_\lambda}\right)}
 +
\frac{\mathcal{D}'\vert S_{\pi\pi KK}\vert\left(m_\rho \Gamma_\rho(s)\cos{(\delta_{\pi\pi KK})-(m_\rho^2-s)\sin{(\delta_{\pi\pi KK})}}\right)}{\left(1+\frac{s}{\Lambda^2_{\lambda'}}\right)}
\nonumber \\
&+&
\frac{\mathcal{E}m_\rho\Gamma_\rho(s)}{1+\frac{s} {\Lambda^2_{\lambda}}}
+\frac{\mathcal{E}^\prime\vert S_{\pi\pi KK}\vert\left\{(m^2_\rho-s)\cos[\delta_{\pi\pi KK
}(s)]+m_\rho\Gamma_\rho(s)\sin[\delta_{\pi\pi KK}(s)]\right\}}{1+\frac{s}{
\Lambda^2_{\lambda^\prime}}}
\nonumber \\
&+&   \mathcal{F}\left((m_\rho^2-s)Re[\hat{f}^0_0(s)]+m_\rho\Gamma_\rho(s) Im[\hat{f}^0_0(s)]\right)
   +\mathcal{G}\left((m_\rho^2-s)Im[\hat{f}^0_0(s)]-m_\rho\Gamma_\rho(s) Re[\hat{f}^0_0(s)]\right) \Bigg]
\nonumber \\ 
&+&   
\frac{\mathcal{H}Re[\hat{f}^0_0(s)]}{1+\frac{s}{\Lambda^2_{\lambda}}} 
+\frac{\mathcal{H}^\prime\vert S_{\pi\pi KK}\vert\left\{Im[\hat{f}^0_0(s)][\cos[\delta_{\pi\pi KK}(s)]- Re[\hat{f}^0_0(s)]\sin[\delta_{\pi\pi KK}(s)]\right\}}{1+\frac{s}{\Lambda^2_{\lambda^\prime}
}}  
\nonumber \\ 
&+&
\frac{\mathcal{P}Im[\hat{f}^0_0(s)]}{1+\frac{s} {\Lambda^2_{\lambda}}} 
+  \frac{\mathcal{P}^\prime\vert S_{\pi\pi KK}\vert\left\{Re[\hat{f}^0_0(s)]\cos[\delta_{\pi\pi KK
}(s)]+Im[\hat{f}^0_0(s)]\sin[\delta_{\pi\pi KK}(s)]\right\}}{1+\frac{s}{\Lambda^2_{
\lambda^\prime}}} + \mathcal{Q}|\hat{f}^0_0(s)|^2,
\label{eq:missing} 
\end{eqnarray}
\end{widetext}
where the whole  scalar partial wave $\hat f^0_{\pi\pi\to\pi\pi}$ is approximated, as usual, only with its isoscalar part:
\begin{equation}
    \hat{f}^0_0(s)=\frac{\eta^0_0(s) 
    e^{2i \delta^0_0(s)}-1}{2i},
\end{equation}
with the elasticity parameter $\eta^0_0(s)$ and phase shift $\delta^0_0(s)$ taken from the ``global parameterization" proposed in \cite{Pelaez:2019eqa} up to 2 GeV (Note that in that work $\hat f$ is called $\hat t$). It satisfies $S^0_0(s)=1-2i \hat f^0_0$, as defined in the main text. Let us recall that Ref.\cite{Pelaez:2019eqa} provides three different parameterizations between 1.4 and 2 GeV,  due to the existence of three incompatible data sets.
Here we have shown the results for  parameterization I,  since the other two yield rather similar curves differing slightly only above 1.5 GeV, but still describing the data with a dominant $\pi\pi\to K\bar K$ contribution (although with somewhat different $\mathcal{ A, B...}$ parameters).

%

This extended model for $\Delta\Gamma_{\pi\pi}$ contains additional contributions and parameters. However, the net contribution of terms
containing $\pi\pi \to KK$ FSI effects,  i.e. those with primed or hat parameters, still dominate very strongly the asymmetry in the 1 to 1.5 GeV region.
This can be observed in Fig.\ref{fig:plotKpipi}, where we show as a dashed line the sum of the $\mathcal{\hat B, D^\prime, E^\prime, H^\prime}$ and $\mathcal{P^\prime}$ terms, whereas the remaining contribution 
corresponds to the dotted curve. Note that all other terms are also strongly constrained by data below the 1 GeV region. Therefore, the main conclusion of this work remains the same:  
$\pi\pi \to KK$ rescattering effects dominate the $s$ dependence of the CP asymmetry in the region between 1 and 1.5~GeV.

For completeness we provide below the central values of the parameters for this extended model:
\begin{eqnarray}
 {\cal A}=-7.76,&\quad&  {\cal \hat B}=25.35, \quad  \Phi=85.55^{\rm o},\nonumber\\
{\cal C} = 0.77,&\quad& {\cal D}=-22.92, \quad  {\cal D'}=28.62, \nonumber \\
 {\cal E}=-19.53, &\quad& {\cal E'}=-5.95, \quad  {\cal F}=17.53, \nonumber \\
 {\cal G}=-11.38,  &\quad& {\cal H}=-4.61, \quad {\cal H'}=-22.91,\nonumber\\
 {\cal P}=3.54, &\quad&{\cal P'}=-0.22, \quad {\cal Q}=3.46 . \nonumber
 \end{eqnarray}


\begin{thebibliography}{40}
\bibitem{footnote1}{This may imply an additional change of sign 
when using the $S^{1/2}$-matrix formalism \cite{Cheng:2016shb}, that may serve to alleviate the finding that the $\pi\pi\leftrightarrow K\bar K$ interaction was headed in the wrong direction in that model.}
\end{thebibliography}

%

\end{document}